# Doping evolution of the gap structure and spin-fluctuation pairing in Ba(Fe$_{1-x}$Co$_x$)$_2$As$_2$ superconductors

A. E. Karakozov[1], M. V. Magnitskaya[1,2], L.S. Kadyrov[3], and B. P. Gorshunov[3*]

[1] *L.F. Vereshchagin Institute for High Pressure Physics, Russian Academy of Sciences, 108840 Troitsk, Moscow, Russia*
[2] *P.N. Lebedev Physical Institute, Russian Academy of Sciences, 119991 Moscow, Russia*
[3] *Moscow Institute of Physics and Technology (State University), 141700 Dolgoprudny, Moscow Region, Russia*

*bpgorshunov@gmail.com

Doping dependence of the superconducting state structure and spin-fluctuation pairing mechanism in the Ba(Fe$_{1-x}$Co$_x$)$_2$As$_2$ family is studied. BCS-like analysis of experimental data shows that in the overdoped regime, away from the AFM transition, the spin-fluctuation interaction between the electron and hole gaps is weak, and Ba(Fe$_{1-x}$Co$_x$)$_2$As$_2$ is characterized by three essentially different gaps. In the three-gap state an anisotropic (nodeless) electron gap $\Delta_e(x,\varphi)$ has an intermediate value between the dominant inner $\Delta_{2h}(x)$ and outer $\Delta_{1h}(x)$ hole gaps. Close to the AFM transition the electron gap $\Delta_e(x, \varphi)$ increases sharply and becomes closer in magnitude to the dominant inner hole gap $\Delta_{2h}(x)$. The same two-gap state with close electron and inner hole gaps $\Delta_{2h}(x) \approx \Delta_e(x, \varphi)$ is also preserved in the phase of coexisting antiferromagnetism and superconductivity. The doping dependence of the electron gap $\Delta_e(x, \varphi)$ is associated with the strong doping dependence of the spin-fluctuation interaction in the AFM transition region. In contrast to the electron gap $\Delta_e(x, \varphi)$, the doping dependence of the hole gaps $\Delta_{1,2h}(x)$ and the critical temperature $T_c(x)$, both before and after the AFM transition, are associated with a change of the density of states $\gamma_{nh}(x)$ and the intraband electron-phonon interaction in the hole bands. The non-phonon spin-fluctuation interaction in the hole bands in the entire Co concentration range is small compared with the intraband electron-phonon interaction and is not dominant in the Ba(Fe$_{1-x}$Co$_x$)$_2$As$_2$ family.

The high-$T_c$ iron-based superconductors (FeSCs) are multiband quasi-two-dimensional compounds with strongly anisotropic Fermi surface and low carrier density in the hole-like and electron-like bands [1]. The Fermi surface (FS) of these compounds consists of hole-like (h) pockets at the $\Gamma$ point and electron-like (e) pockets centered at the X = ($\pi$, 0) and Y = (0, $\pi$) points of the Brillouin zone. Compared to strongly correlated high-$T_c$ cuprates, which are similar in their basic characteristics, electron-electron correlations in FeSCs are not large (see, for example, reviews [2, 3]).The parent orthorhombic (Ort) Fe-based compounds are antiferromagnetic (AFM) metals of spin-density wave (SDW) type with the magnetic ordering vectors **Q** = ($\pi$, 0), (0, $\pi$). Unlike dielectric parent high-$T_c$ cuprates, they have free electronic states at the FS that are not associated with magnetism but can, in principle, be involved in superconducting (SC) pairing. The electronic structure of these compounds is very sensitive to small changes in doping, pressure, and degree of disorder. When in parent compounds the magnetic atoms Fe (3d$^6$) in the *a–b* plane are replaced by atoms with larger number of *d* electrons (electron doping) or the non-magnetic atoms out of this plane are replaced by atoms with smaller valence (hole doping), antiferromagnetism is gradually suppressed which leads to the onset of superconductivity. In this regime, the AFM and SC gaps coexist at the Fermi surface (the coexistence of antiferromagnetism and superconductivity or underdoped regime). Maximal critical temperature of SC transition $T_c$ is reached at the total suppression of magnetism (the optimal regime). Further increase in doping (overdoped regime) results in a reduction of $T_c$ down to the total suppression of superconductivity. Close to the optimal doping, an orthorhombic-to-tetragonal (Ort–Tet) phase transition occurs in the system. In the hole-doped FeSCs, the structural Ort–Tet and AFM–nonmagnetic transitions occur simultaneously. The electron doping, when Fe is substituted with Co, Ni, etc., promotes the isotropization of the spin/orbital order in the *a*–*b* plane and an electronic transition to a state analogous to the nematic phase in liquid crystals, which precedes the structural transition. Nematic fluctuations and the associated softening of the shear modulus and Ort structural distortion $a \ne b$ reaches a maximum near the Ort–Tet transition and decreases in the overdoped phase as $T_c$ decreases [4, 5]. The gap structure of FeSCs depends significantly on the composition, the quality of the samples, and the external parameters. According to various experiments, SC order parameter can have the s-wave as well as d-wave (with nodes on some FS areas) symmetry [6].

A strong anisotropy of the SC order parameter is usually associated with a non-phonon (electron-electron) mechanism of superconductivity. Although electron-electron correlations in Fe-based compounds are relatively small, the electron-electron mechanisms of SC pairing associated with the observed in FeSCs s-wave spin fluctuations for large wave vectors $\mathbf{q} \approx \mathbf{Q}$ (inter-pocket) [7, 8] and quadrupole $d_{x^2-y^2}$-wave charge fluctuations for small wave vectors $\mathbf{q}$ (intra-pocket) [9, 10], may be important for the superconductivity of these compounds [11–15]. Also under discussion is the relationship of superconductivity with the electronic nematic fluctuations [4, 5].

The pairing mechanisms mediated by the spin and orbital (charge) fluctuations are considered in the spin-fluctuation theory of superconductivity of FeSCs [11–24]. According to the spin-fluctuation theory (see, for example, review [18]), in FeSCs the basic non-phonon pairing channel is the inter-pocket electron–hole interaction $V_{eh}(\mathbf{k}_e - \mathbf{k}_h' \approx \mathbf{Q}) = V_{eh}$. Because of the proximity to the AFM phase, this channel is enhanced by spin fluctuations with the same wave vector $\mathbf{Q}$ and always exceeds intra-pocket Coulomb repulsion $|V_{eh}| > |V_e; V_h|$ in FeSCs with hole and electron pockets. Depending on the sign of $V_{eh}$, either a sign-reversed gap on electron and hole FSs (for $V_{eh} > 0$, this is the so-called $s_\pm$ state), or conventional $s_{++}$-state ($V_{eh} < 0$) is formed in the system. Schematic phase diagram of gap structure of FeSCs for $V_{eh} > 0$ is shown in Fig. 1.

As is seen in Fig. 1, in the FeSC there is a possible s-wave ($A_{1g}$ symmetry) gap with nodes whose position is determined by the pairing interaction anisotropy and does not contradict symmetry [so-called nodal s-wave symmetry that is usually associated with d-wave ($B_{1g}$) superconductivity, and d-wave ($B_{1g}$ symmetry) gap without nodes]. In the Tet phase (as in Fig. 1), the s-wave order parameter $\Delta(\varphi)$, symmetrical about the diagonal of the Brillouin zone $\Delta(\varphi) = \Delta(\varphi + \pi/2)$, in the representation of leading angular harmonics can be approximated on the hole1,2 FSs (centered at $k = 0$) by the constants $\Delta_{h1,2}(\varphi) = \Delta_{h1,2}$. On the circular (centered at X/Y points) FSs, $\Delta(\varphi) = \Delta_{eX/Y}(\varphi_{X/Y})$ may be written as $\Delta_{eX/Y}(\varphi_{X/Y}) = \Delta_{es} \pm \Delta_{ed}\sqrt{2}\cos(2\varphi_{X/Y})$, with $\Delta_{es}$ the s-wave contribution and $\Delta_{ed}\sqrt{2}\cos 2\varphi_{X/Y}$ the d-wave component of $d_{x^2-y^2}$ symmetry, $\varphi_{X/Y}$ are the angles along electron circular X/Y FSs, measured relative to $k_x$. Depending on the pairing interaction $V(\mathbf{k}, \mathbf{k}')$, the gap anisotropy $\Delta_{ed}/\Delta_{es}$ may be greater than 1 and $\Delta(\varphi)$ will have nodes. In the same approximation, the s-wave BCS pairing interaction $V(\mathbf{k}, \mathbf{k}') = Vu(\mathbf{k})u(\mathbf{k}')$ with $u(\mathbf{k}) =$ const on the hole FSs and $u(\mathbf{k}) = w_s \pm w_d\sqrt{2}\cos 2\varphi_{X/Y}$ on the X/Y FSs may be written as the hole

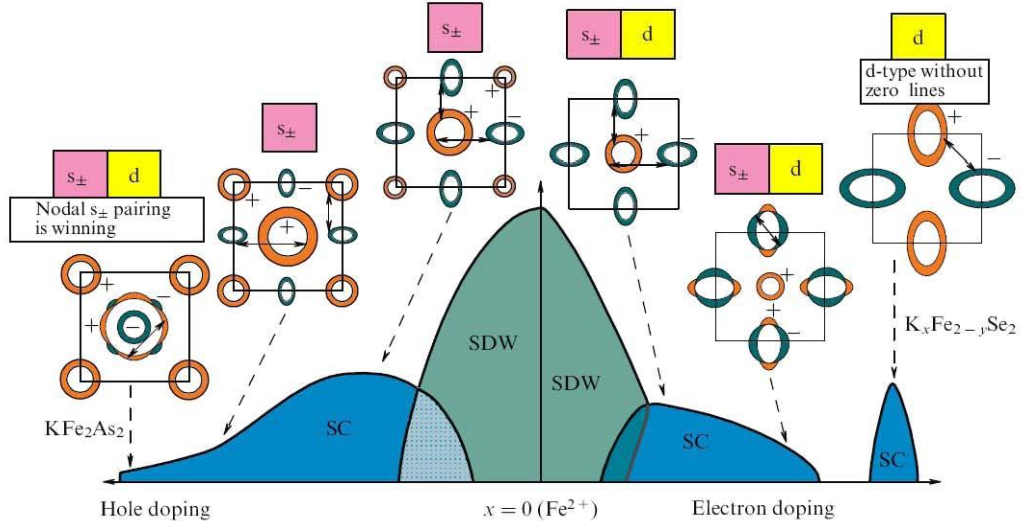

Fig. 1. Schematic phase diagram of iron compounds for both hole and electron dopings. The coexistence of antiferromagnetic (SDW) and superconducting (SC) phases appears on the microscopic level for the case of electron doping, and on the macroscopic level (division into SDW and SC domains) upon hole doping. The qualitative picture of the superconducting parameter symmetries, which follows from the spin-fluctuation theory [20–22] and from the leading angular harmonics approximation (LAHA) [23, 24] for the two-dimensional system is shown on symmetrical Fermi surfaces in the insets above the phase diagram; s and d stand for the predominant and subdominant symmetries of pairing. Solid lines with an arrow at both ends (↔) indicate the predominant interaction. (Courtesy of the *Physics–Uspekhi* journal, see Ref. 19.)

intra-pocket interactions $V_{h1,2}(\varphi,\varphi') = V_{h1,2}$, the electron X/Y intra-pocket interactions

$$V_e u(\varphi_{X/Y}) u(\varphi'_{X/Y}) = V_s(1 \pm k_d \cos 2\varphi_{X/Y})(1 \pm k_d \cos 2\varphi'_{X/Y}), \qquad (1)$$

and the inter-X-Y-pocket interactions

$$U_{XY}(\varphi_X, \varphi'_Y) = U_s(1 + m_d \cos 2\varphi_X)(1 - m_d \cos 2\varphi'_Y), \qquad (2)$$

where $k_d = \sqrt{2}\, w_d/w_s$ (or $m_d$) is the degree of $V_{X/Y}$ ($U_{XY}$) pairing interaction anisotropy. In fact, $k_d^2$ makes sense of the ratio of the (attractive) $V_e w_d^2$ d-component to the repulsive $V_e w_s^2$ s-component of pairing interaction or coupling constants: $k_d^2 = 2\lambda_d/\lambda_s$, where $\lambda_{s,d} = N_e(0) V_e w_{s,d}^2$ and $N_e(0)$ is the density of states at the electron X/Y FSs. In the Ort phase, s + d symmetry (without phase shift between X and Y pockets) of the SC order parameter and the pairing $V_{X/Y}$, $U_{XY}$ interactions is possible. Such symmetry of the SC order parameter may be observed in electron-overdoped FeSCs due to structural Ort $a \neq b$ distortions induced by nematic fluctuations.

The angular dependence of the intra-pocket $V_e(\varphi_{X/Y}, \varphi'_{X/Y})$ interaction minimizes the intra-pocket Coulomb repulsion $V_{es} = V_e w_s^2$ due to the formation of sign-reversed gap on different FS areas. Anisotropy of pairing in the electron band is manifested in the electron-overdoped FeSCs away from the AFM transition and filling the hole bands with doping, when the inter-pocket $V_{eh}$ interaction and the hole gaps $\Delta_{h1,2}$ are small. In this case, the role of the electron band superconductivity increases and it turns out to be an advantageous increase in the X/Y intra-pocket attraction $V_e w^2_d$ and to form a sign-reversed gap with nodes on electron pockets (as in Fig. 1). The degree of $\Delta_e(\varphi)$ anisotropy $\tilde{k}_d = \Delta_{ed}/\Delta_{es}$ in the spin-fluctuation theory is determined by the competition of the pairing in s [mainly by spin inter-pocket interaction $V_{eh}(\mathbf{Q})$] and d (mainly by orbital intrapocket interaction) channels [16, 17]. Electron-phonon attraction reduces repulsion in all non-phonon channels and can, in principle, transform a sign-reversed gap with nodes into a sign-preserved gap with minima. In addition, taking into account the electron-phonon interaction, the influence of the spin-fluctuation mechanism on superconductivity in the electron and hole bands becomes unequal.

Dominant mechanism of superconductivity in various FeSCs and even in various bands is not universal and varies with composition even within a single family [20]. The spin-fluctuation theory is indirectly confirmed by the experimental observation of the spin resonant peak in the spin excitation spectrum [8, 25] (see also [26, 27]). Direct determination of the doping dependence of SC order parameters from experimental data makes it possible to evaluate the role of non-phonon pairing interactions in the superconductivity of various FeSCs.

Direct comparison of theoretical models with experimental data is often rather difficult. In many experiments, one is able to resolve groups (clusters) of SC gaps in two relatively narrow energy ranges, rather than individual gaps in the bands [28]. (Equalization of electron and hole gaps can be a consequence of a strong interband interaction [29]). Analysis of the properties of such "two-gap" superconductors allows one to estimate only the interaction between bands from different clusters. This interaction turns out to be weak, at least, for samples of sufficiently good quality (see reviews [30, 31]). The pairing interactions in the $i$-th and $j$-th bands with the closest gaps (within clusters) cannot be unambiguously determined, since their SC gaps $\Delta_{i,j}(T)$ coincide within the experimental uncertainties. In particular, it is impossible to investigate the basic interactions (which determine the critical temperature $T_c$) in the bands within clusters that combine the maximum gaps. Two-gap superconductors, within the experimental uncertainty, are often equally successfully described by various pairing models, which allow an ambiguous interpretation of SC mechanism in the compounds under study. A more reliable estimate of interaction between the electron and hole SC condensates can be made in superconductors with the significantly different electron $\Delta_e(T)$ and hole $\Delta_h(T)$ SC gaps, in particular, in the "three-gap" FeSCs [32]. The subsequent investigation of the doping evolution of the three-gap state in comparison with the results of spin-fluctuation theory will make it possible to study basic interactions in the FeSCs in more detail.

Here, we present the results of such investigation for the most extensively studied Ba(Fe$_{1-x}$Co$_x$)$_2$As$_2$ family, for which there are relevant experimental data. In this family, the three-gap state occurs in the overdoped compound Ba(Fe$_{0.9}$Co$_{0.1}$)$_2$As$_2$ with $T_c = 20$ K. The SC gaps and interband coupling constants of this compound have been determined [32, 33] on the basis of the BCS-like analysis of terahertz and infrared optical experiments. It was shown that the SC state is characterized by an s-wave SC order parameter with three essentially different gaps: two isotropic hole gaps ($\Delta_{1h} = 15$ cm$^{-1}$, $\Delta_{2h} = 30$–$35$ cm$^{-1}$) and an anisotropic (nodeless) electron gap $\Delta_e(\varphi)$ with an amplitude of 21 cm$^{-1}$ and $V_e(\varphi, \varphi')$ pairing interaction anisotropy $k_d = 0.5$ (see Eq. 1), and a weak h2-e interband interaction.

The three-gap structure of the SC state in Ba(Fe$_{0.9}$Co$_{0.1}$)$_2$As$_2$ determined in [32] is fully confirmed by the measurements of reflection coefficient [34] and ARPES [35]. According to the ARPES data, the gaps with $\Delta_{1h} \approx 15$ cm$^{-1}$ and $\Delta_{2h} \approx 35$ cm$^{-1}$ refer to the outer and inner hole bands, respectively. The smallness of the interaction of the electron and hole bands in Ba(Fe$_{0.9}$Co$_{0.1}$)$_2$As$_2$ is clearly confirmed by the non-BCS temperature dependence of the superfluid density $\rho_s(t = T/T_c)$ that reveals a pronounced region of inflection at intermediate temperatures $t$ [32, 33]. Such dependence is characteristic of two-gap superconductors with a weak interaction of the bands that belong to different clusters [36]. In the three-gap superconductors, such a dependence also indicates a weak interaction of the bands with intermediate and largest gap values [the electron and inner hole bands in Ba(Fe$_{0.9}$Co$_{0.1}$)$_2$As$_2$]. Studying the evolution with decreasing Co content of the three-gap overdoped state in Ba(Fe$_{1-x}$Co$_x$)$_2$As$_2$ with a weak inter-pocket interaction V$_{eh}$ to an underdoped regime when antiferromagnetism and superconductivity coexist, allows one to provide a deeper understanding of basic interactions in this family.

A systematic experimental study of superconductivity in the Ba(Fe$_{1-x}$Co$_x$)$_2$As$_2$ compounds within entire range of Co concentrations, $x$, was carried out by Hardy et al. [37]. The authors measured the electronic heat capacity $C_s(x, t)$ and determined the density of states $\gamma_n(x)$ at the FS which are available for SC pairing. According to [37], the doping dependence $\gamma_n(x)$ in Ba(Fe$_{1-x}$Co$_x$)$_2$As$_2$ exhibits a maximum near the optimal doping and correlates with the concentration dependence of the critical temperature $T_c(x)$.

Comparison of temperature dependences of normalized electronic heat capacity $c_s(x, t) = C_s(x, T)/\gamma_n(x)T$ for the compounds with close $\gamma_n(x)$ and $T_c(x)$ values in the overdoped and underdoped regimes (see Fig. 2a) provides information on behavior of the intermediate e-gap $\Delta_e(x)$ with decreasing doping. In the underdoped compound, the normalized heat capacity $c_s(x, t) = C_s(x, T)/\gamma_n(x)T$ decreases significantly (by up to 20–25%) at intermediate temperatures and increases near $T_c$, in accordance with the entropy conservation: $\int_0^1 c_s(x,t)dt = 1$. A similar difference in the behavior of temperature dependence $c_s(x, t)$ in the overdoped and underdoped regimes is even more clearly expressed in some other FeSCs families, for example, the 111-compounds NaFe$_{1-x}$Co$_x$As [38] (see Fig. 2b). Within the two-band α-model [39], this is formally explained by a pronounced redistribution of the density of states from the cluster of small-value energy gaps to the cluster with large gap values (see Fig. 2e in [37]). In the three-gap compounds, in particular Ba(Fe$_{1-x}$Co$_x$)$_2$As$_2$, this behavior of $c_s(x, t)$ can be explained by a dramatic equalization of the intermediate electron $\Delta_e$ and dominant hole $\Delta_{2h}$ gaps in the coexistence regime under a smooth change in the density of states in bands $\gamma_{nj}(x)$. Such equalization of the gap values can be a consequence of an increase in the interband interaction (see, e.g. [29]).

To calculate the heat capacity $c_s(x, t)$ of three-band superconductors, we used the BCS-like equations [32, 45], which are a correct generalization of the multiband BCS equations [40, 41] for the case of strong coupling in the spirit of the α-model [42].

The coupling strength is usually characterized by the deviation of the ratio $2\Delta(0)/T_c$ from the BCS value $2\Delta(0)/T_c^0 = 2\alpha_0 = 3.52$. Using an empirical recipe known as the α-model [42], the BCS approach successfully describes the properties of conventional superconductors with strong coupling. Strong coupling effects reduce $T_c$ due to the smearing of the SC gap and an increase in the number of quasiparticles compared to that in the BCS model [43]. In calculating the temperature dependences of SC gaps in the bands, $\Delta_j(T)$, this circumstance can be taken into account by redefining the quasiparticle distribution functions in the BCS equations, with the condition that the calculated critical temperature is equal to the experimental value of $T_c$. For a single band, e.g., this condition is satisfied by the distribution function of the form $f[(T_c/T_c^0)\sqrt{\varepsilon^2 + \Delta^2(T)}/T] = f[\alpha_0\sqrt{\omega^2 + \delta^2(t)}/t]$, where $\varepsilon$ is the quasiparticle energy, $\omega = \varepsilon/\Delta(0)$, $\delta(t) = \Delta(t)/\Delta(0)$ is the reduced gap and $f$ is the Fermi function. The BCS-like equation with such distribution function ensures the equality of the calculated and experimental $T_c$, the equality of the α-parameter to the experimental value $\alpha = \Delta(0)/T_c$, and has the solution $\delta(t) = \delta_0(t)$ as assumed in the α-model.

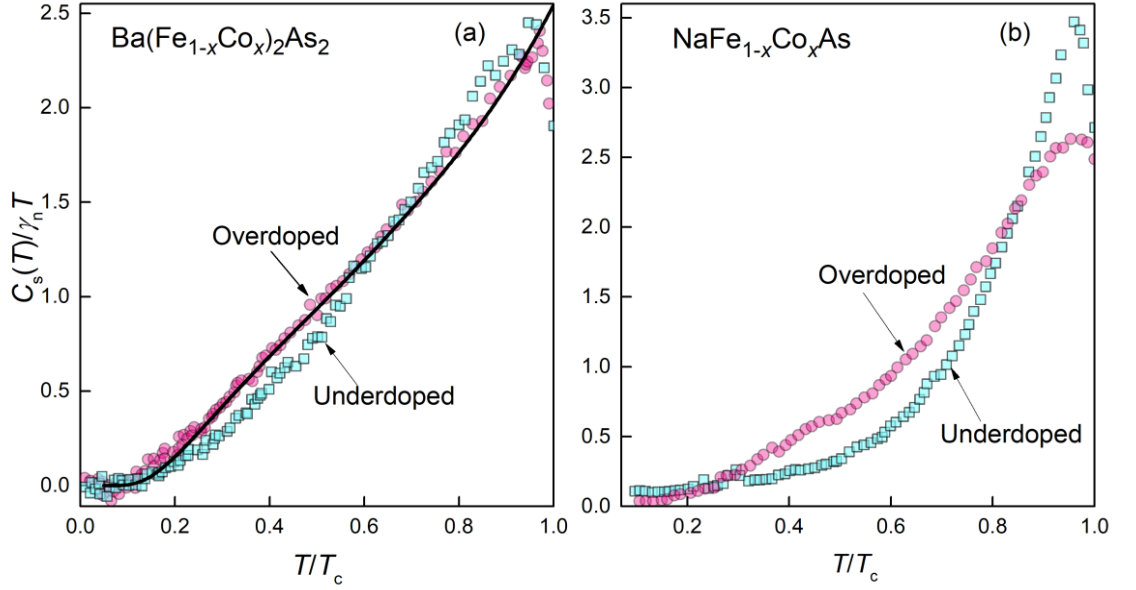

Fig. 2. The normalized temperature dependence of the electronic heat capacity $C_s(T)$ for compounds with close values of $T_c$ and $\gamma_n$. (a) Overdoped Ba(Fe$_{0.91}$Co$_{0.09}$)$_2$As$_2$, $T_c = 20.7$ K, $\gamma_n = 16.4$ mJ mol$^{-1}$ K$^{-2}$ (circles) and underdoped Ba(Fe$_{0.95}$Co$_{0.05}$)$_2$As$_2$, $T_c = 19.5$ K, $\gamma_n = 14.7$ mJ mol$^{-1}$ K$^{-2}$ (squares); adapted from [37]. The solid line shows $C_s(T)$ for the overdoped three-gap compound Ba(Fe$_{0.9}$Co$_{0.1}$)$_2$As$_2$ ($T_c \approx 20$ K) calculated with parameters taken from [32]. (b) Overdoped NaFe$_{0.95}$Co$_{0.05}$As, $T_c \approx 18.1$ K, $\gamma_n = 6.3$ mJ mol$^{-1}$ K$^{-2}$ (circles) and underdoped NaFe$_{0.975}$Co$_{0.025}$As, $T_c \approx 20.1$ K, $\gamma_n = 7.6$ mJ mol$^{-1}$ K$^{-2}$ (squares); adapted from [38].

The formal application of the α-model prescription for the two-band superconductor, $\delta_1(t) = \delta_2(t) = \delta_0(t = T/T_c)$, known as the two-band α-model [39], well fits the behavior of the heat capacity $c_s(t)$ with any number of bands and in some cases allows one to determine the gaps $\Delta_{1,2}(0)$ quite accurately (e.g., for MgB$_2$, see [44]), but it is absolutely unsuitable for analysis of characteristics that are more sensitive to the temperature dependence of the gaps $\delta_{1,2}(t)$ such as tunnel spectra and superfluid density $\rho_s(t)$ [32, 33]. This model does not take into account the fact that the gaps $\delta_{1,2}(t)$ depend on the interband interactions and the strong-coupling corrections in the bands are not the same (in the band with the larger gap $\Delta_2$ the correction is always greater than in the band with a smaller gap $\Delta_1$). A more correct generalization of the α-model is possible on the basis of the two-band BCS equations [40, 41] for reduced gaps $\delta_{1,2}(t) = \Delta_{1,2}(t)/\Delta_{1,2}(0)$:

$$\ln \delta_1(t) = -n_1(t) - \Lambda_{12}\{1 - \delta_2(t)/\delta_1(t)\}, \qquad (3)$$

$$\ln \delta_2(t) = -n_2(t) - \Lambda_{21}\{1 - \delta_1(t)/\delta_2(t)\}, \qquad (4)$$

$$n_J(t) = 2\int_0^\infty d\omega \, f[\alpha_J \varepsilon_J(\omega)/t]/\varepsilon_J(\omega) \,, \tag{5}$$

$$\varepsilon_J(\omega) = \sqrt{\omega^2 + \delta_J^2(t)} \,. \tag{6}$$

Here, $\alpha_J = \Delta_J(0)/T_c^0$, $n_{1,2}(t)$ is the quasiparticle density in the bands 1,2. The interband terms proportional to the constants $\Lambda_{12}$ and $\Lambda_{21}$ have opposite signs and describe the transfer of pairs from condensate with a large gap $\Delta_2$ to a condensate with a smaller gap $\Delta_1$. As a result, the gaps $\delta_1(t)$ and $\delta_2(t)$ approach each other [45]. The interband constants in (3), (4)

$$\Lambda_{12} = \lambda_{12}/\theta(0), \Lambda_{21} = \lambda_{21}\theta(0), \quad \theta(0) = \Delta_1(0)/\Delta_2(0) \tag{7}$$

depend on the effective coupling constants $\lambda_{12}$ and $\lambda_{21}$, which are a combination of all the bare coupling constants of the usual BCS-type form $\lambda_{IJ}^0 = V_{IJ} N_J(0)$:

$$\lambda_{IJ} = \lambda_{IJ}^0/D^0 \,, \quad D^0 = \lambda_{11}^0 \lambda_{22}^0 - \lambda_{12}^0 \lambda_{21}^0 \,. \tag{8}$$

In contrast to $\lambda_{IJ}^0$, the effective constants $\lambda_{IJ}$ can reach large values even for $\lambda_{IJ}^0 \ll 1$ (see examples in [45]).

For a correct generalization of the BCS equations (3) – (6), it is necessary to take into account the standard renormalization of the bare BCS constants $\lambda_{IJ}^0 \to \bar{\lambda}_{IJ} = \lambda_{IJ}^0/(1 + \lambda_{II}^0 + \lambda_{I\neq J}^0)$ ($<1$) and replace the effective constants (7), (8) by their renormalized values $\lambda_{IJ} \to \tilde{\lambda}_{IJ}$, $\Lambda_{IJ} \to \tilde{\Lambda}_{IJ}$ and also in the spirit of the α-model, use in (5) the distribution function of a more general form $f[\tilde{\alpha}_j \sqrt{\omega^2 + \delta_j^2(t)}/t]$, where $\tilde{\alpha}_{1,2}$ are determined by the self-consistency equation, which follows from the equality of the critical temperatures in the bands

$$\ln\frac{\tilde{\alpha}_2}{\alpha_0} = \frac{\tilde{\Lambda}_{21}}{\tilde{\Lambda}_{12} + \ln\frac{\alpha_0}{\tilde{\alpha}_1}} \ln\frac{\alpha_0}{\tilde{\alpha}_1} \tag{9}$$

and the fitting parameter $\tilde{\alpha}_1$ in the interval from $\Delta_1(0)/T_c$ (weak coupling in the band 1, $\Delta_1 \ll \Delta_2$) to $\alpha_0$ (strong coupling, $\Delta_1 \sim \Delta_2$) [32, 45]. The parameter $\tilde{\alpha}_1$ can be approximately determined from the interpolation relation

$$\tilde{\alpha}_1 \approx \Delta_1(0)/T_c + (\alpha_0 - \Delta_1(0)/T_c)\, \Delta_1(0)/\Delta_2(0), \tag{10}$$

where the ratio of the SC gap values is taken as a measure of coupling strength. In this case, the number of fitting parameters in the BCS and BCS-like equations is the same.

The equations (3), (4) are invariant with respect to the change in the sign of the interband interaction $\lambda_{I\neq J}^0 \to -\lambda_{I\neq J}^0$ and the sign of one of the gaps $\Delta_J$ and coincide for $s_\pm$ and $s_{++}$ pairing.

For superconductors with known characteristic frequencies of pairing interactions $\Omega_{\log}$ in the region of a relatively weak interband interaction, the BCS-like analysis of experimental data makes it possible to determine all interaction constants with sufficient accuracy using a minimal number of fitting parameters. In particular, all the gaps $\Delta_J(t)$ and the intraband and interband coupling constants $\lambda_{IJ}^0$ we found from both the tunneling spectra of the $Mg_{1-x}Al_xB_2$ system [46] and the temperature dependence of the electronic heat capacity $c_s(t)$ of MgB2 [45] coincide with the results of first-principles calculations.

As the interband interactions increase, the determinant of the BCS-like system $\overline{D} = \overline{\lambda}_{11}\overline{\lambda}_{22} - \overline{\lambda}_{12}\overline{\lambda}_{21}$ decreases, the effective coupling constants $\tilde{\Lambda}_{IJ}$ increase sharply, and the gaps $\delta_1(t)$ and $\delta_2(t)$ become closer in magnitude. In a special case, when the determinant $\overline{D}$ vanishes, the solutions of the system (3) – (6) are linearly dependent:

$$\Delta_1(t)/\Delta_2(t) = \overline{\lambda}_{12}/\overline{\lambda}_{22} = \overline{\lambda}_{11}/\overline{\lambda}_{21} , \qquad (11)$$

and the gaps $\delta_1(t) = \delta_2(t)$ coincide [47, 45]. In this region, small (within the experimental uncertainty) changes of $\delta_1(t) \approx \delta_2(t)$ lead to strong changes in effective constants $\tilde{\Lambda}_{12}, \tilde{\Lambda}_{21}$ and large uncertainty in the definition of bare coupling constants $\lambda_{IJ}^0$, which makes the analysis of the experimental data (both in the BCS-like and Eliashberg approaches) ambiguous and allows, in general, different pairing scenarios. (Note that in the region of strong interband interaction $\delta_1(t) \approx \delta_2(t)$ and the use of the two-band α-model is completely justified.)

Multi-band BCS-like equations can be simplified, given the smallness of some interband interactions. In particular, in the three-band superconductor, for example, in $Ba(Fe_{0.9}Co_{0.1})_2As_2$ with the largest gap $\Delta_{2h}$, the interband constants $\tilde{\Lambda}_{2h(e,1h)}$ are much smaller than in the other two bands,

$$\tilde{\Lambda}_{2h(e,1h)}\big/\tilde{\Lambda}_{e,1h} = (\gamma_{e,1h}\big/\gamma_{2h})(\Delta_{e,1h}\big/\Delta_{2h})^2 \ll 1 , \qquad (12)$$

and they can be neglected by assuming $\delta_{2h}(t) \approx \delta_0(t)$ and taking into account only the most significant interband interactions $\tilde{\Lambda}_{1h2h}$ and $\tilde{\Lambda}_{e2h}$ of the 1h and e bands, with the 2h band that determines $T_c$ [32]. For other Co concentrations, due to the possible increase in the interaction of the electron and inner hole bands in the region where antiferromagnetism and superconductivity coexist, it is necessary to take into account the interaction $\tilde{\Lambda}_{2he}$, too.

The BCS-like equations for the reduced gaps $\delta_j(t) = \Delta_j(t)/\Delta_j(0)$ for hole gaps $j$ = 1h, 2h have the form:

$$\ln \delta_{2h}(t) = -\tilde{n}_{2h}(t) - \tilde{\Lambda}_{2h,e}\left\{1 - \delta_e(t)\big/\delta_{2h}(t)\right\} , \qquad (13)$$

$$\ln \delta_{1h}(t) = -\tilde{n}_{1h}(t) - \tilde{\Lambda}_{1h2h}\left\{1 - \delta_{2h}(t)\big/\delta_{1h}(t)\right\} , \qquad (14)$$

$$\tilde{n}_j(t) = 2\int_0^\infty d\omega\, f[\tilde{\alpha}_j \varepsilon_j(\omega)/t]/\varepsilon_j(\omega) , \qquad (15)$$

$$\varepsilon_j(\omega) = \sqrt{\omega^2 + \delta_j^2(t)} , \qquad (16)$$

where $\tilde{n}_j(t)$ is the quasiparticle density and $\varepsilon_j(\omega)$ is the reduced spectrum of the j-th band.

In the equation for the electron gap in Ba(Fe$_{1-x}$Co$_x$)$_2$As$_2$, it is necessary to take into account the anisotropy of the intra-X/Y-pocket $V_{X/Y}(\varphi_{X/Y}, \varphi'_{X/Y}) = V_s(1 \pm k_d\cos2\varphi_{X/Y})(1 \pm k_d\cos2\varphi'_{X/Y})$ and inter-XY-pocket $U_{XY}(\varphi_X, \varphi'_Y) = U_s(1 + m_d\cos2\varphi_X)(1 - m_d\cos2\varphi'_Y)$ pairing in the electron band. The X/Y gaps on the electron FS $\Delta_{eX/Y}(\varphi_{X/Y}) = \Delta_{es}[1 \pm \tilde{k}_d\cos(2\varphi_{X/Y})]$ and pairing interactions $V_{X/Y}$, $U_{XY}$ differ only in $\pi/2$ phase shift. The equation for the amplitude $\Delta_{es}$ and the anisotropy degree $\tilde{k}_d$ of the X/Y gaps do not depend on the phase shift and are the same for X and Y pockets in both Tet and Ort phases, so in such calculations we can confine ourselves to the case of the Ort s + d symmetry (with zero phase shifts) of the $V_{X/Y}$, $U_{XY}$. It can be shown that the pairing interaction $V_e$ in this equation is the sum of intra-pocket $V$ and inter-pocket $U$ pairing interactions. For the same anisotropy $k_d = m_d$, this interaction can be written in a simple form $V_e(\varphi, \varphi') = \lambda_{es}\gamma_{ne}u(\varphi)u(\varphi')$, with $u(\varphi) = 1 + k_d\cos2\varphi$, $\lambda_{es} = (V + U)/\gamma_{ne}$. When interacting with the isotropic 2h-band, the s-component of the electron gap $\Delta_{es}$ is renormalized due to the transfer of s-pairs from 2h-band, the gap anisotropy $\tilde{k}_d$ does not coincide with the initial $k_d$ and depends on temperature: $\tilde{k}_d(t) = \bar{\lambda}_{es}(\tilde{\lambda}_{2h} - \tilde{\lambda}_{e2h}\Delta_{2h}(t)/\Delta_{es}(t))k_d$. [In the special case of $\bar{D} = 0$, the function $\tilde{k}_d(t)$ is finite (11)]. By implication, the gap anisotropy $\tilde{k}_d$ differs from the pairing anisotropy $k_d$ in the ratio of the gap $\Delta_0(t)$ (without $V_{e2h}$ interaction) to the gap $\Delta_{es}(t)$: $\tilde{k}_d/k_d \approx \Delta_0(t)/\Delta_{es}(t) < 1$. For weak interband interaction $\tilde{k}_d \approx 1$, for the gaps inside the cluster $\Delta_{es}(t) \approx \Delta_{2h}(t)$, $\tilde{k}_d$ does not depend on temperature and is $\tilde{k}_d \approx k_d\Delta_{es}(0)/\Delta_0(0)$.

For the electron gap $\Delta_e(\varphi, t) = \Delta_{es}(0)\delta_e(t)\beta(\varphi, t)$, where $\beta(\varphi, t) = 1 + \tilde{k}_d(t)\cos(2\varphi)$, the reduced gap $\delta_e(t)$ is determined by the BCS-like equation:

$$\ln \delta_e(t) + \langle \bar{\beta}(\varphi,t)\ln\beta(\varphi,t) - \bar{\beta}(\varphi,0)\ln\beta(\varphi,0)\rangle =$$
$$-2\tilde{n}_e(t) + \frac{\tilde{\Lambda}_{e2h}}{\langle u_\varphi\rangle\langle\beta(\varphi,0)\rangle}\left\{\frac{\delta_{2h}(t)\langle\beta(\varphi,0)\rangle}{\delta_e(t)\langle\beta(\varphi,t)\rangle} - 1\right\}, \quad (17)$$

$$\tilde{n}_e(t) = \left\langle \bar{\beta}(\varphi,t)\int_0^\infty d\omega\, \frac{f(\tilde{\alpha}_e\varepsilon_e(\omega,\varphi,t)/t)}{\varepsilon_e(\omega,\varphi,t)}\right\rangle, \quad (18)$$

$$\varepsilon_e(\omega,\varphi,t) = \sqrt{\omega^2 + \delta_e^2(t)\beta^2(\varphi,t)}, \quad (19)$$

with the averages $\langle F\rangle \Rightarrow \frac{1}{2\pi}\left\langle\int_0^{2\pi} u_\varphi F d\varphi\right\rangle$ and $\bar{\beta}(\varphi,t) = \beta(\varphi,t)/\langle\beta(\varphi,t)\rangle$.

The solutions of the system of equations (13), (14), (17) are used to calculate the normalized electron entropy $s_j(x, t)$ of the bands and the heat capacity $c_s(x, t)$:

$$s_j(t) = 3\frac{\alpha_j}{\pi^2}\int_{-\infty}^\infty d\omega \int_0^{2\pi}\frac{d\varphi}{2\pi}\{[\alpha_j\varepsilon_j(\omega,\varphi)/t]f(\alpha_j\varepsilon_j(\omega,\varphi)/t) -$$
$$\ln[f(-\alpha_j\varepsilon_j(\omega,\varphi)/t)]\}. \quad (20)$$

Here, $\alpha_j$ is equal to the experimental value, $\alpha_j = \Delta_j(0)/T_c$, $\varepsilon_j$ are the spectra (14), (17), and

$$c_s(t) = \frac{d}{dt}\left[\gamma_{1h} s_{1h}(t) + \gamma_e s_e(t) + \gamma_{2h} s_{2h}(t)\right], \qquad (21)$$

where $\gamma_j = \gamma_{nj}/\gamma_n$ is the fractional density of states ($\gamma_{1h} + \gamma_e + \gamma_{2h} = 1$).

To analyze the doping evolution of the gap structure and spin-fluctuation pairing, we used the relevant experimental data on the $c_s(x, t)$ for high-quality samples of the Ba(Fe$_{1-x}$Co$_x$)$_2$As$_2$ family at $0.1 < x < 0.4$ [37] for $t < 0.95$ outside the region of thermodynamic fluctuations near $T_c$. Figures 2a and 3a–d show the results of our calculations of $c_s(t)$ within the model (13), (14), (17), which differs from [32], where $\lambda_{2he} = 0$, $\tilde{\alpha}_{2h} = \alpha_0$, $\delta_{2h}(t) = \delta_0(t)$.

In Figure 2a, the solid line shows the heat capacity $c_s(t)$ of the three-gap Fe-based superconductor Ba(Fe$_{0.9}$Co$_{0.1}$)$_2$As$_2$ ($T_c = 20$ K), with SC gaps and symmetry $\Delta_e(\varphi, t = 0)$ defined in [32], a weak 2h-e interaction $\tilde{\lambda}_{2he} = 0.12$ ($\tilde{\lambda}_{e2h} = 0.45$, $\tilde{\lambda}_{1h2h} = 0.4$), and fractional density of states $\gamma_e = 0.2$, $\gamma_{2h} = 0.5$ ($\gamma_{1h} = 0.3$), in comparison with the experimental dependence $c_s(t)$ of the close compound Ba(Fe$_{0.91}$Co$_{0.09}$)$_2$As$_2$ ($T_c = 20.7$ K). The difference between the interband constants and [32] is mainly due to a rather large experimental uncertainty of optical measurements. Despite this difference, this result basically confirms the three-gap structure with the dominant intraband pairing in the 2h-band in Ba(Fe$_{0.9}$Co$_{0.1}$)$_2$As$_2$.

The doping dependence of the gap structure $\Delta_j(x)$ and $T_c(x)$ is related to a change in the intraband pairing $\lambda^0_{II} \approx V(q \to 0)\gamma_{ni}(x)$ and interband interaction $\lambda^0_{I \neq J} \approx V(Q, x)\gamma_{nj}(x)$. $V(Q, x)$ associated with the strong doping dependence of the spin fluctuation intensity [50], increases with decreasing electron doping to reach a maximum near the AFM transition. As the doping decreases, the density of states $\gamma_{nj}(x)$ increases in the hole bands and decreases in the electron band.

For small changes in the doping $\delta x = x - x_0$, a good initial approximation for superconductor with dominant intraband pairing is given by $\Delta_j(x,0)/T_c = \Delta_j(x_0,0)/T_c(x_0)$, with interband constants $\tilde{\lambda}(x) = \tilde{\lambda}(x_0)$ and gap anisotropy $\tilde{k}_d(x) = \tilde{k}_d(x_0)$, which does not change the reduced gaps $\delta_j(x, t) = \delta_j(x_0, t)$ [see Eqs. (13), (14), and (17)]. In this approximation, the doping dependence of the heat capacity $c_s(x, t)$ [see Eqs. (20) and (21)] is associated only with a change in the fractional density of states $\gamma_j(x)$ (and, respectively, $\gamma_{nj}(x)$ and the intraband pairing constants) in the bands. Figure 3a shows the temperature dependence of the heat capacity $c_s(t)$ for the compound Ba(Fe$_{0.925}$Co$_{0.075}$)$_2$As$_2$ ($T_c = 22.9$ K) with the same interband constants, parameters $\tilde{\alpha}_j$, anisotropy $\tilde{k}_d \approx k_d = 0.5$ and $\gamma_{1h} = 0.3$, $\gamma_e = 0.15$, $\gamma_{2h} = 0.55$ ($\gamma_n = 19.2$ mJ mol$^{-1}$ K$^{-2}$), as in Ba(Fe$_{0.9}$Co$_{0.1}$)$_2$As$_2$. The variations in $\Delta_j$ and interband coupling constants relative to the initial configuration that are allowed by experimental uncertainty do not exceed 5%. An increase in $\gamma_{nj}(x)$ in the hole bands and a decrease in the electron band correspond to the dependence of the $\gamma_{nj}(x)$ electron and hole bands in the FeSCs with decreasing electronic doping.

For the optimally doped compound Ba(Fe$_{0.9425}$Co$_{0.0575}$)$_2$As$_2$ with $T_c = 24.3$ K, the heat capacity calculated in the initial approximation agrees well with the experimental dependence $c_s(t)$ only for a very small value $\gamma_e \leq 0.03$ ($\gamma_n = 18.7$ mJ mol$^{-1}$ K$^{-2}$ [37]), which actually corresponds to a dramatic equalization of the electron $\Delta_{es}(0)$ and hole $\Delta_{2h}(0)$ gaps (see Fig. 2 and corresponding discussion in the text). An analysis of the temperature dependence $c_s(t)$ shows that the electron $\Delta_{es}(0)$ and inner hole $\Delta_{2h}(0)$ gaps that satisfy the condition for a smooth change in $\gamma_{ne}$ ($0.1 \leq \gamma_e \leq 0.15$) should have a value in the range 39 cm$^{-1}$ to 43 cm$^{-1}$ and the gap $\Delta_e(\varphi)$ anisotropy $\Delta_{es}\tilde{k}_d$ of the order of the e–2h cluster size. The gaps in the cluster coincide with experimental accuracy $\Delta_{es}(t) \approx \Delta_{2h}(t)$, and, as a result, the interband e–2h contribution to the system (13), (17) is small and the reduced gaps $\delta_{2h}(t) \approx \delta_e(t)$ weakly depend on the interband e–2h interactions. Within the limits of experimental uncertainty, superconductivity of optimally doped $Ba(Fe_{0.9425}Co_{0.0575})_2As_2$ and other similar two-gap superconductors can be formally explained by both conventional phonon and non-phonon (for example, in the optimally hole-doped $Ba_{0.68}K_{0.32}Fe_2As_2$ model [49]) scenarios. The study of the doping dependence of the SC gap structure in the Ba(Fe$_{1-x}$Co$_x$)$_2$As$_2$ family in comparison with the results of the spin-fluctuation theory allows one to clarify the role of the non-phonon mechanism in superconductivity of optimally doped Ba(Fe$_{0.9425}$Co$_{0.0575}$)$_2$As$_2$ and compounds in the phase of coexisting antiferromagnetism and superconductivity.

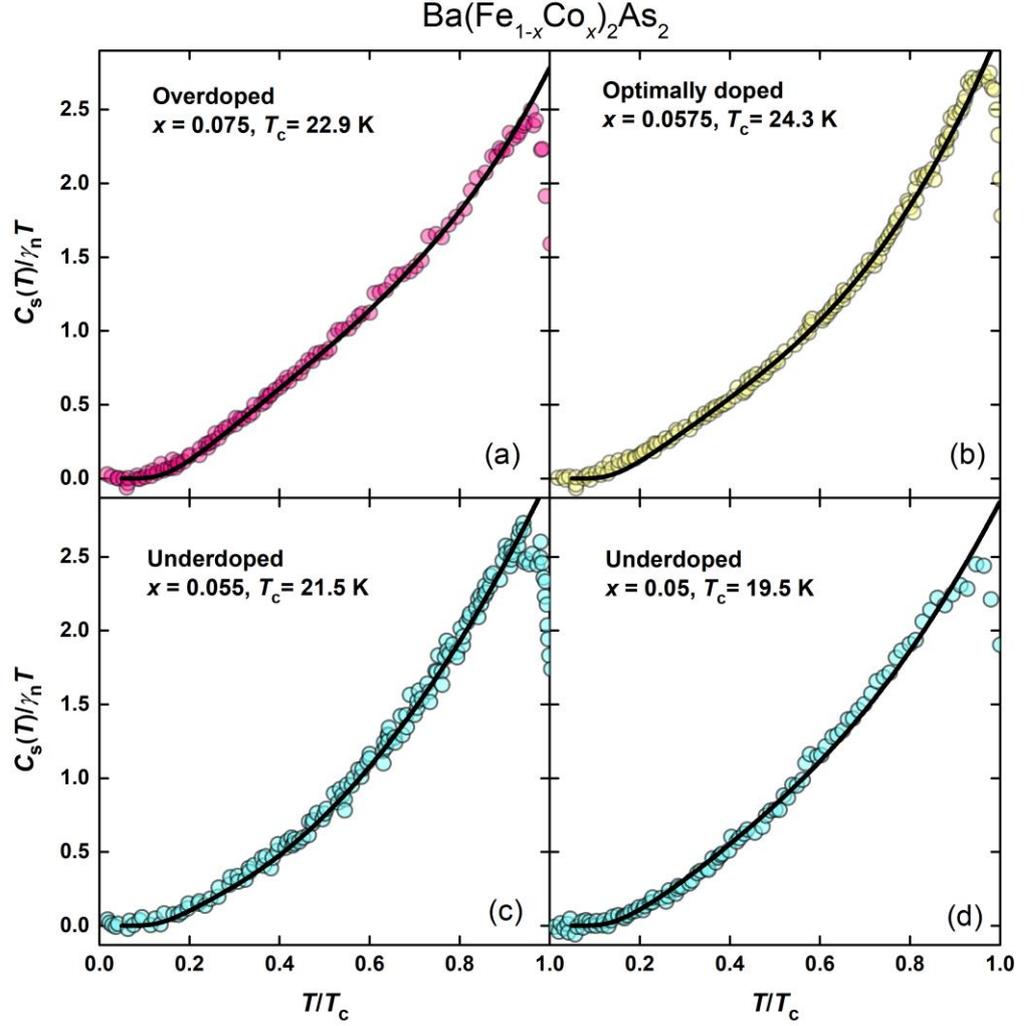

Fig. 3. The normalized electronic heat capacity $C_s(T)$ for $Ba(Fe_{1-x}Co_x)_2As_2$. (a) The three-gap overdoped compound $Ba(Fe_{0.925}Co_{0.075})_2As_2$ with $x = 0.075$, $T_c = 22.9$ K; (b) The optimally doped compound with $x = 0.0575$, $T_c = 24.3$ K; (c, d) The overdoped compounds with $x = 0.055$, $T_c = 21.5$ K and $x = 0.05$, $T_c = 19.5$ K, correspondingly. Symbols show the experimental data from [31], the lines indicate the results of approximation according to equations (13), (14), (17).

Away from the AFM transition, a basic non-phonon mechanism is possible in superconductors with strongly anisotropic FS (one-band high-$T_c$ cuprates, K$_x$Fe$_{2-y}$Se$_2$, and multiband FeSCs with a dominant electron gap $\Delta_{es}(0) > \Delta_{1,2h}(0)$ [see Fig. 1 and Eq. (12)]. The s-wave superconductivity of overdoped Ba(Fe$_{1-x}$Co$_x$)$_2$As$_2$ away from the AFM transition with dominant intraband pairing and weak interaction with the correlated electron band cannot be explained by spin-fluctuation theory with repulsive intraband interaction, without taking into account the strong electron-phonon interaction. In these compounds, the correlation effects in the isotropic hole bands are small as $\tilde{\lambda}_{2he}$, and clearly manifest themselves only in the anisotropy of the gap $\Delta_e(\varphi)$ in the electron band, for which the intraband pairing is enhanced by interaction with d$_{x2-y2}$ charge (orbital) fluctuations with small wave vectors q. Within the limits of error, the e-2h interaction and the e-gap anisotropy do not change with decreasing doping. The doping dependence of the critical temperature of the overdoped Ba(Fe$_{1-x}$Co$_x$)$_2$As$_2$ away from the AFM transition can be explained by an increase in conventional intraband pairing in the 2h band due to an increase in the density of states $\gamma_{n2h}(x)$. Close to the AFM transition, correlation effects reach a maximum. A dramatic equalization of the electron $\Delta_{es}$ and $\Delta_{2h}$ hole gaps and the decrease in the electron gap $\Delta_e(\varphi)$ anisotropy in the optimally doped Ba(Fe$_{1-x}$Co$_x$)$_2$As$_2$ can be explained by a sharp increase in the e-2h interband interaction with the spin fluctuations accompanying the AFM transition, as is assumed by the spin-fluctuation theory [48]. This scenario is illustrated in Fig. 3b that shows the heat capacity $c_s(t)$ of Ba(Fe$_{0.9425}$Co$_{0.0575}$)$_2$As$_2$ with close 2h and e-bands ($\Delta_{2h}$ = 43 cm$^{-1}$ and $\Delta_{es}$ = 39 cm$^{-1}$), strong e-2h interband interaction $\tilde{\lambda}_{e2h}$ = 2.0, $\gamma_{2h}$ = 0.59, $\gamma_e$ = 0.12, and e-gap anisotropy $\tilde{k}_d \approx k_d\Delta_{es}(t)/\Delta_{es}(x = 0.1, t) = 0.1$. Superconductivity in the anisotropic e-band of this compound is mainly determined by the spin-fluctuation mechanism; however, a small increase by ≈ 5% in $\Delta_{2h}$ and $T_c$ during the transition to the optimal regime, shows that this mechanism in Ba(Fe$_{1-x}$Co$_x$)$_2$As$_2$ is not dominant.

The same two-gap state with close electron and inner hole gaps is also preserved in the phase of coexisting antiferromagnetism and superconductivity. For the two-gap state, it is possible to determine only the characteristics of the outer hole band, $\Delta_{1h}(x)$ and $\gamma_{1h}(x)$, the average characteristics of the e-2h cluster $\Delta_{2h}(x) \approx \Delta_e(x)$, $\gamma(x) = \gamma_{2h}(x) + \gamma_e(x)$, and the interaction of the outer hole band with a cluster $\tilde{\lambda}_{1h2h}(x)$. The doping evolution of the SC state structure for underdoped Ba(Fe$_{1-x}$Co$_x$)$_2$As$_2$ is well described by the two-gap initial approximation $\Delta_{1,2h}(x,0)/T_c$ = $\Delta_{1,2h}(x_0,0)/T_c(x_0)$, $\tilde{\lambda}_{1h2h}(x) = 0.3$ with $\gamma_{1h}(x)$=0.3, $\gamma(x) = 0.7$ (Figs. 3c and 3d).

Over the entire range of Co concentrations $0.1 > x > 0.4$, the calculated strong coupling parameters $\alpha_{1,2h} = \Delta_{1,2h}(x)/T_c(x)$ and fractional density of the states $\gamma_{1h}(x)$ and $\gamma(x)$, within the calculation error of ~5% does not depend on doping [$2\Delta_{1h}(x)/T_c(x) = 2$, $2\Delta_{2h}(x)/T_c(x) = 5$, $\gamma_{1h}(x) = 0.3$, and $\gamma(x) = 0.7$] and the doping dependence of the density of states in the 1,2 hole bands $\gamma_{nh}(x)$ with an accuracy of $\gamma_e(x)/\gamma_{2h}(x) \ll 1$ is proportional to the total density of states $\gamma_n(x)$ in Ba(Fe$_{1-x}$Co$_x$)$_2$As$_2$.

Figure 4 shows schematically the doping dependence of the SC gap structure of Ba(Fe$_{1-x}$Co$_x$)$_2$As$_2$ calculated on the basis of experimental data and the doping dependence of the density of states $\gamma_n(x)$ for Ba(Fe$_{1-x}$Co$_x$)$_2$As$_2$ experimentally determined by Hardy et al. [37].

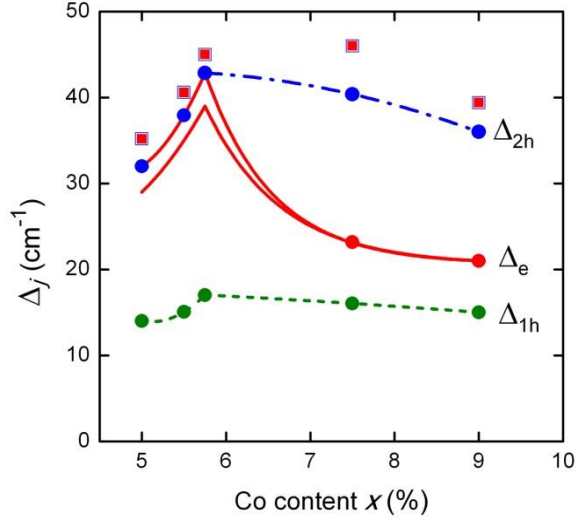

Fig. 4. Evolution of the superconducting gaps in Ba(Fe$_{1-x}$Co$_x$)$_2$As$_2$ with the doping level $x$. Calculated values are shown by circles. The region of the e-2h cluster is between the solid lines. Squares indicate the experimental density of states $\gamma_n(x)$ available for the SC pairing (arb. units) [37].

Variation of the isotropic hole gaps over the entire range of Co concentrations correlates with a change in the density of states in the bands $\gamma_{nh}(x)$ [~$\gamma_n(x)$] and can be explained by the doping dependence of the intraband electron-phonon interaction. The complex doping dependence of the magnitude and anisotropy of the e-gap $\Delta_e(x, \varphi)$ is mainly determined by the doping dependence of the spin-fluctuation mechanism [48]. Away from the AFM transition in the overdoped regime with a relatively weak interband e-2h interaction, Ba(Fe$_{1-x}$Co$_x$)$_2$As$_2$ is in the three-gap state with e-gap that is significantly smaller than the dominant 2h-gap. Close to the AFM transition in the optimally doped phase and in the phase of coexisting antiferromagnetism and superconductivity, due to the increase in the e-2h interaction, the electron gap $\Delta_e(x)$ sharply increases approaching the dominant internal hole gap $\Delta_{2h}(x)$, and Ba(Fe$_{1-x}$Co$_x$)$_2$As$_2$ goes into a two-gap state with close $\Delta_{es}(x)$ and $\Delta_{2h}(x)$ gaps. Strong e-2h interaction with the isotropic 2h-band increases the s-component of the electron gap $\Delta_e(x, \varphi)$ and significantly reduces the e-gap anisotropy $\tilde{k}_d$ in this region. In the phase of coexisting antiferromagnetism and superconductivity, the hole gaps and the electron gap coupled with the inner hole gap by the strong e-2h interaction decrease with decreasing doping, as does the density of states available for the SC pairing due to competition with an opening AFM gap. The solid lines in Fig. 4 outline the range of possible electron gap values in the e-2h cluster. (When constructing the graph, we took into account that the intermediate electron gap $\Delta_e(x)$ cannot exceed the dominant inner hole gap $\Delta_{2h}(x)$ only at the expense of increasing the e-2h interband interaction [29].)

An analysis of the doping dependence of the gap structure in Ba(Fe$_{1-x}$Co$_x$)$_2$As$_2$ shows that the evolution of the critical temperature $T_c(x)/\Delta_{1,2h}(x) \approx$ const is not associated with the strong dependence of the spin-fluctuation interaction $V(Q, x)$ on doping and can be explained by the doping dependence of the density of states $\gamma_{2h}(x)$ and the electron-phonon interaction in the inner hole band with the dominant gap $\Delta_{2h}(x)$. The non-phonon spin-fluctuation mechanism significantly affects superconductivity in the electron band with anisotropic FS, but the spin-fluctuation interaction in the inner hole band in the entire Co concentration range is small compared with the intraband electron-phonon interaction and has a weak effect on $T_c$ in the Ba(Fe$_{1-x}$Co$_x$)$_2$As$_2$ family.

The work was supported in part by Russian Foundation for Basic Research (Projects ##18-02-01075, 16-02-01122, and 19-02-00509) and the Russian Ministry of Education and Science (Program 5-100). We are grateful to S.A. Kuzmichev for fruitful scientific discussions and comments.

———————————————————————